\documentclass[aps,prb,reprint,groupedaddress]{revtex4-1}
\usepackage{graphicx}  

\begin{document}


\title{Observation of Incompressibility at $\nu=4/11$ and $\nu=5/13$ }

\author{N. Samkharadze$^1$, I. Arnold$^1$, 
L.N. Pfeiffer$^2$, K.W. West$^2$, 
        and G.A. Cs\'{a}thy$^{1,3}$ \footnote{gcsathy@purdue.edu}}

\affiliation{${}^1$ Department of Physics and Astronomy, Purdue University, West Lafayette, IN 47907, USA \\
${}^2$Department of Electrical Engineering, Princeton University, Princeton, NJ 08544\\
${}^3$ Birck Nanotechnology Center Purdue University, West Lafayette, IN 47907, USA \\}

\date{\today}

\begin{abstract}

The region of filling factors $1/3<\nu<2/5$
is predicted to support new types of fractional quantum Hall states with topological order
different from that of the Laughlin-Jain or the Moore-Read states. Incompressibility is a necessary 
condition for the formation of such novel topological states. 
We find that at 6.9~mK incompressibility develops
only at $\nu=4/11$ and $5/13$, while the states at $\nu=6/17$ and $3/8$ remain compressible.
Our observations at $\nu=4/11$ and $5/13$
are first steps towards understanding emergent topological order in these fractional quantum Hall states.

\end{abstract}

\maketitle

The study of electronic systems with topological properties is one of the most active areas of research in 
contemporary condensed matter physics. This field of study was opened up
by the discovery of the integer quantum Hall states developing in the two-dimensional electron gas (2DEG)
subjected to a perpendicular magnetic field ($B$) \cite{qh}.
Topological phases distinct from the IQHSs
also exist at zero magnetic field in topological insulators and topological superconductors \cite{topoinsul}.
We witnessed a rapid development of the theory of topological order evident
in efforts to classify topological phases, to identify topological invariants, as well as to extend  
the theory beyond the known topological phases \cite{wenN}.

Strong interactions in many-body systems are recognized 
as being instrumental in the generation of emergent topological order.
A textbook example of such emergent order is that of
the conventional fractional quantum Hall states (FQHS)
described by Laughlin's wavefunction and Jain's theory of composite fermions (CF) \cite{tsui,Laughlin,Jain}.
However, it is well known that electron-electron interactions also lead to FQHSs with 
order fundamentally different from that of the conventional Laughlin-Jain states.
One example of such an exotic FQHS is the even denominator FQHS at $\nu=5/2$,
which is thought to be described by a wavefunction of the Moore-Read type \cite{willett,wen,moore,nayak}.

The observation of a depression, i.e. a local minimum, in the magnetoresistance $R_{xx}$ 
at the filling factor $\nu=4/11$ \cite{WPan03}
gave a new impetus to the quest for an enlarged family of topological ground states. 
Early theoretical work suggested that, owing to the residual interactions between the CFs,
at $\nu=4/11$  the ground state is an unusual FQHS \cite{Sitko}. In contrast, subsequent work
found that this state may be either a conventional FQHS \cite {goerbig,park,chang2,chang} or even an electron crystal \cite{lee1,lee2,shibata04,shibata05}. 
Work by W\'ojs, Yi, Quinn and collaborators reinforced a novel FQHS at $\nu=4/11$, termed
the WYQ state \cite{AWojs04,AWojs05,AWojs06,AWojs07,AWojs09b}. 
Most recently  the idea of novel topological order at $\nu=4/11$ of the WYQ type received strong support
from CF diagonalization over an extended Hilbert space \cite{mukherjee14}
and which included a more accurate model of the interaction of CFs than the earlier used one  \cite{AWojs04,AWojs05}. 
Even though a trial wavefunction for the WYQ state remains yet to be constructed, 
calculations reveal that the WYQ state is topologically distinct \cite{AWojs04,AWojs05,mukherjee14} 
from either the Laughlin-Jain \cite{Laughlin,Jain} or the Moore-Read state \cite{moore}. 
This difference is quantified by a different shift, a topological invatiant of the Haldane sphere \cite{Haldane,wen92}.

Incompressibility is a necessary condition for establishing topological order.
While the theoretical support favoring a
FQHSs with a novel topological order at $\nu=4/11$ 
is compelling \cite{AWojs04,AWojs05,mukherjee14}, 
an unambiguous experimental confirmation of incompressibility of the 
ground states at this filling factor is still missing. 
Incompressibility is signaled by the opening of an energy gap $\Delta$ in the density of states which in
transport measurements manifests itself in an activated magnetoresistance of the form
$R_{xx} \propto \exp ( - \Delta / 2 k_B T )$. Here  $T$ is the temperature and $k_B$ Boltzmann's constant.
Testing for an activated behavior is of importance since
a depression in $R_{xx}$ may not always develop into an activated $R_{xx}$, hence
such a depression does not guarantee the development of a FQHS.
Examples of filling factors at which a fractional quantum Hall ground state does not develop
in spite of the presence of a depression in $R_{xx}$
are $\nu=1/7$ \cite{goldman88, WPan02} and $2/11$ \cite{WPan02}.

In this manuscript we present a study of correlated states in the $1/3<\nu<2/5$ range at ultra-low temperatures, a factor of 5 below that 
of previous studies at these filling factors \cite{WPan03}. We report an opening of an energy gap at $\nu=4/11$ as seen in
the observation of an activated magnetoresistance. 
Furthermore, the temperature dependence of magnetotransport at $\nu=5/13$ is also consistent with the development of an incipient gap.
Our results  establish a fractional quantum Hall ground state at these two filling factors and
open up the possibility for existence of novel topological order of the WYQ type in this region.
However, despite a considerable progress at $\nu=4/11$ and $5/13$, 
the transport data in our sample at other filling factors of interest, 
such as $\nu=3/8$ and $6/17$, do not exhibit signatures of an energy gap.
The nature of the ground state at these two filling factors remains uncertain.

\begin{figure}
 \includegraphics[width=1\columnwidth]{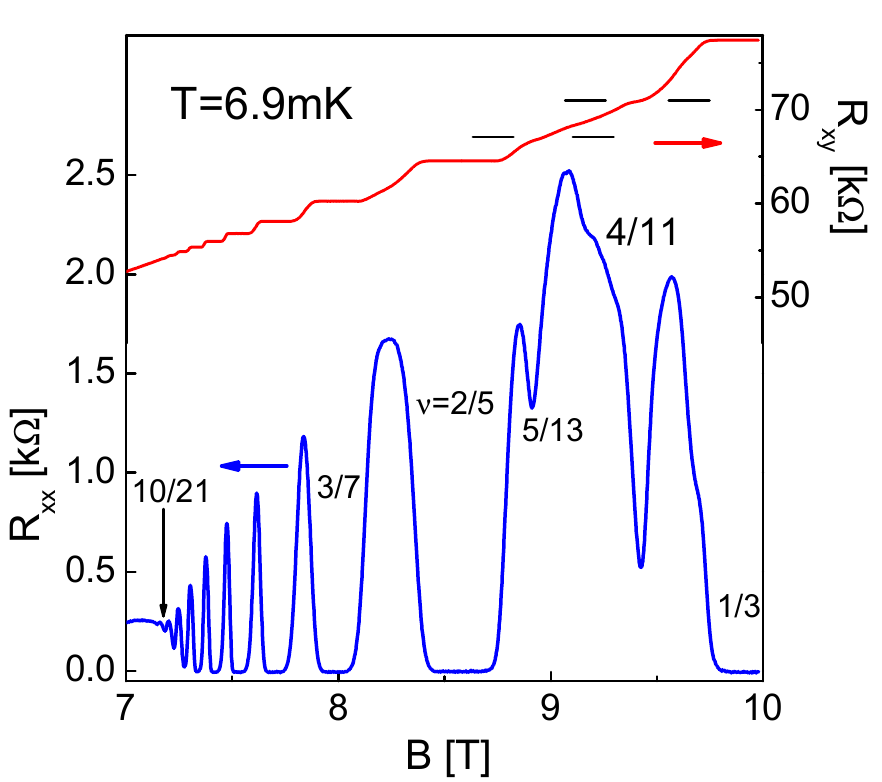}
 \caption{\label{f1}
 Longitudinal ($R_{xx}$) and Hall ($R_{xy}$) resistance in the filling factor range $1/3<\nu<1/2$ at 6.9~mK. 
 Horizontal lines indicate the expected plateau-like behavior of the Hall resistance at $\nu=4/11$ and $\nu=5/13$.
}
\end{figure}

The sample used in this study is a symmetrically doped $56$~nm wide GaAs/AlGaAs quantum well with electron density $8.3 \times 10^{10}$ cm$^{-2}$ 
and mobility $12 \times 10^{6}$ cm$^{2}$/Vs. When cooling the sample, it is of the essence to ensure that the electronic temperature follows that
of the phonons. We achieved this in our experiment by the use of a He-3 immersion cell \cite{WPan04,cell}. Electrons are thermalized
by being passed through sintered silver heat exchangers directly attached onto the sample and which are immersed into the He-3 bath.
The temperature of the He-3 bath is monitored via a quartz tuning fork thermometer, which provides magnetic field independent
readings to the lowest temperatures \cite{cell}.

Fig.1 shows the longintudinal magnetoresistance $R_{xx}$ and the Hall resistance $R_{xy}$ in the region of interest $1/3 < \nu <1/2$
at 6.9~mK. A long series of conventional FQHSs  \cite{Jain} are observed at $\nu=n/(2n+1)$ with $n$ being an integer
up to 10, demonstrating a sample of high quality. Between the two strong conventional FQHSs at $\nu=1/3$ and $2/5$ we notice 
the presence of several features. For example, at $\nu=4/11$ and $5/13$ we observe deep minima in the magnetoresistance and
strong plateau-like inflections in the Hall resistance at $11e^2/4h$ and $13e^2/5h$, respectively. These features
indicate the possibility of FQHSs at these filling factors.

\begin{figure}
 \includegraphics[width=0.8\columnwidth]{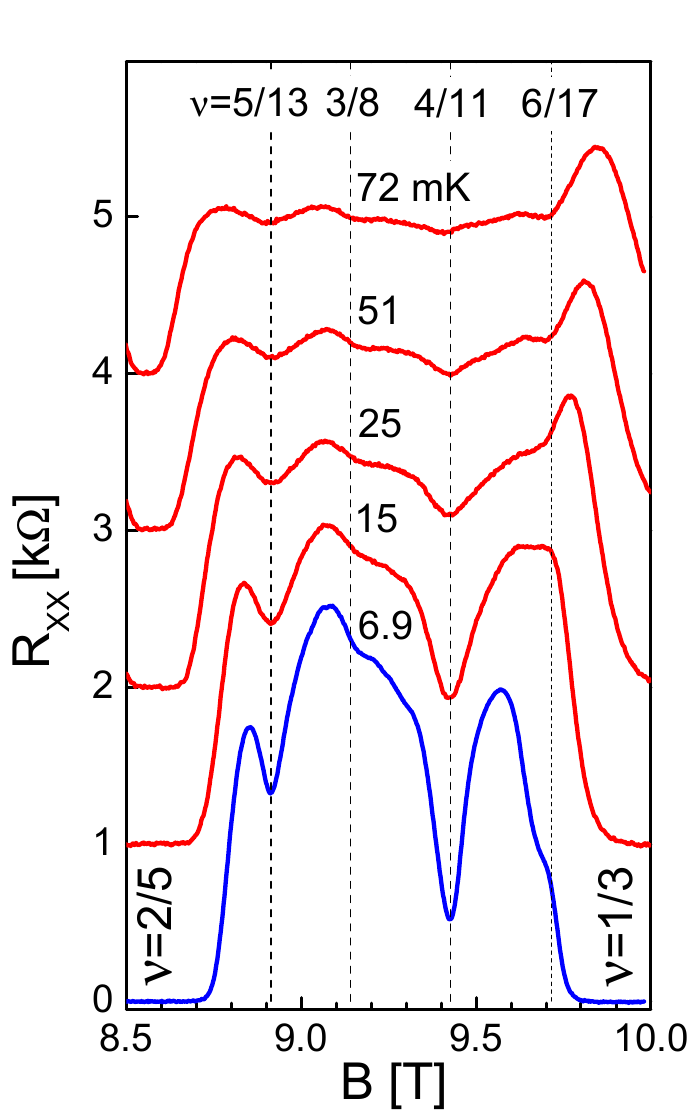}
 \caption{\label{f2}
 Waterfall plot of magnetoresistance curves at several different temperatures in the $1/3<\nu<2/5$ range.
 The dashed lines mark $\nu=5/13$, $3/8$, $4/11$ and $6/17$ filling factors. For clarity, traces are shifted vertically
 by 1~k$\Omega$ each relative to the trace of lower temperature.
}
\end{figure}

A better appreciation of the features of $R_{xx}$ can be gained from Fig.2  by examining the region  of interest $1/3 < \nu < 2/5$
on a magnified horizontal scale. 
At the relatively high temperature of $72$~mK, there are four discernible depressions in the magnetoresistance curve at 
$\nu=5/13$, $3/8$, $4/11$ and $6/17$ filling factors. These depressions are quite similar to those
seen in Ref.\cite{WPan03} at 35~mK.
However, as the temperature is decreased not all of these depressions evolve similarly.
At $\nu=4/11$, $5/13$, and $3/8$ the depressions persist as $T$ is lowered.
In fact, the depressions at $\nu=4/11$ and $5/13$ clearly get more pronounced at $6.9$~mK. 
In contrast, the feature at $\nu=6/17$ develops quite differently from those at the other three filling factors.
As $T$ is lowered, this depression significantly
weakens at $25$~mK and it has virtually disappeared at 15~mK.
At the lowest achieved temperature of 6.9~mK, the $R_{xx}$ shows only a weak inflection near $\nu=6/17$, which is not separated from the $\nu=1/3$ plateau with a local maximum.
Similarly, $R_{xy}$ does not show any hints of quantization.
This suggests that the system at the electronic filling factor $\nu=6/17$  tends to localize in  the limit of zero temperature, joining the $\nu=1/3$ plateau.
Thus we conclude that our data is not consistent with earlier an report of a fractional quantum Hall ground state at  $\nu=6/17$ \cite{WPan03}.

\begin{figure}[t]
 \includegraphics[width=1\columnwidth]{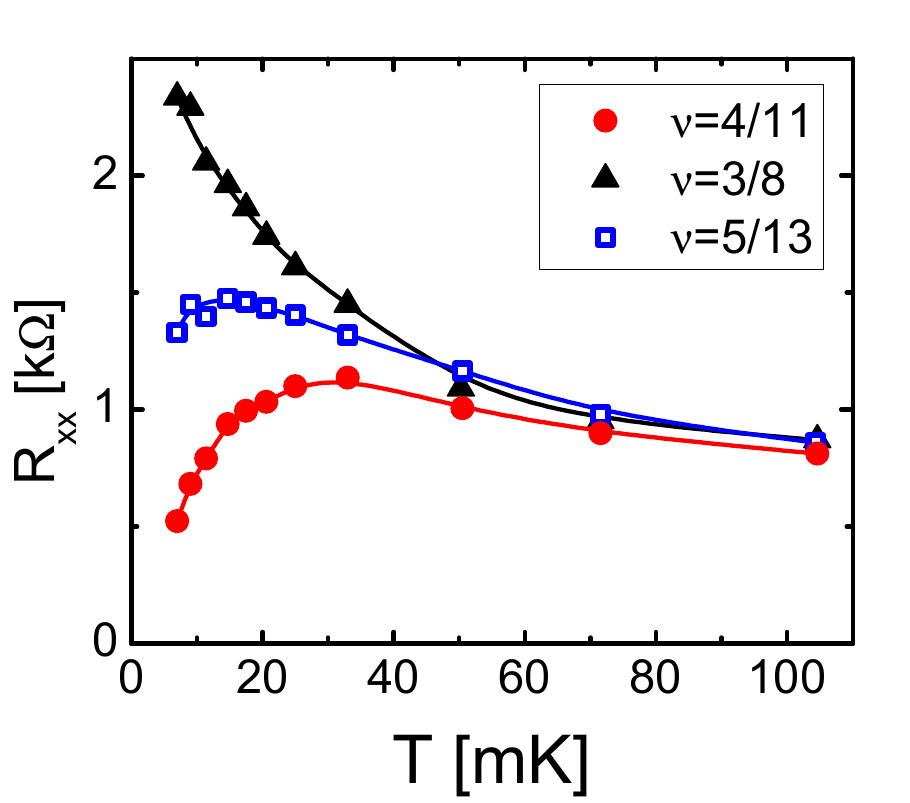}
 \caption{\label{f3}
 Dependence of magnetoresistance on temperature at the filling factors $\nu=4/11, 3/8$ and $5/13$. Lines are guides to the eye.
}
\end{figure}

Further insight on the ground states at $\nu=4/11$, $5/13$, and $3/8$ emerges from Fig.3, where
$R_{xx}$ is plotted against $T$. Since significant differences between $R_{xx}$ at these three
filling factors develop only below $T \approx 30$~mK, Fig.3 demonstrates that
accessing these low temperatures is a critical capability in the study of the $1/3<\nu<2/5$ region.
We notice that as the temperature is lowered from 100~mK, $R_{xx}$ at $\nu=4/11$ 
first increases, then it drops significantly. These data are consistent with 
a vanishing $R_{xx}$ in the $T=0$ limit. Similarly, at the lowest temperatures, $R_{xx}$ at $\nu=5/13$ is
also decreasing, albeit this decrease is not as pronounced as that at $\nu=4/11$.  The observed behavior at
$\nu=4/11$ and $5/13$ indicates incompressibility and, therefore, it
establishes the ground state at these two filling factors is a genuine FQHS. 

An examination of the temperature dependence of the magnetoresistance at $\nu=4/11$ below $T=30$~mK reveals that it is activated.
Indeed, as shown in the Arrhenius plot Fig.4, at the lowest temperatures we find a linear dependence of
$\ln R_{xx}$ on $1/T$. The slope of this linear portion of the data yields an energy gap of $\Delta_{4/11}=15$~mK.
A similar analysis of the incipient incompressibility at $\nu=5/13$ results in an upper bound of the energy gap
$\Delta_{5/13} \simeq 3$~mK.

\begin{figure}[b]
 \includegraphics[width=1\columnwidth]{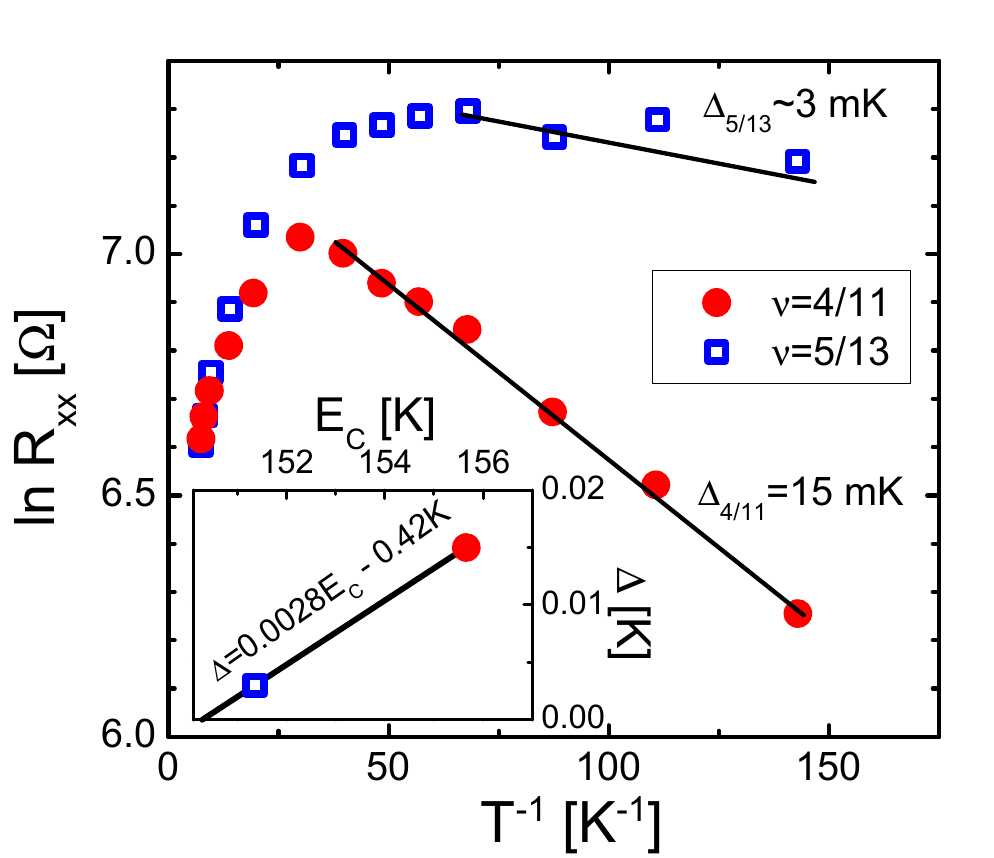}
 \caption{\label{f4}
 Arrhenius plot for $R_{xx}$ at $\nu=4/11$ and $\nu=5/13$ filling factors. Straight lines are used as guides for the eye representing the activated
 regions. The inset shows the  scaling of the experimentally extracted gaps $\Delta$ at $\nu=4/11$ and $5/13$ states with the Coulomb energy $E_C$. 
}
\end{figure}

The FQHS at $\nu=4/11$ is quite feeble as seen from its small energy gap $\Delta_{4/11}=15$~mK. 
This value comes short by more then one order of magnitude when compared to the expected value
$0.002E_C \approx 0.3$~K  from recent  numerical work \cite{mukherjee14}.
Here $E_C=e^2/4 \pi \epsilon l_B$ is the Coulomb energy, $l_B=\sqrt{\hbar/eB}$ is the magnetic length, 
and $\epsilon$ is the dielectric constant in GaAs.  
However, in Ref.\cite{mukherjee14} disorder effects are not included.
In order to estimate the disorder-free, or intrinsic gap $\delta_{4/11}$ of the $\nu=4/11$ FQHS, 
we employ a method of scaling of the gaps of a pair of particle-hole conjugate FQHSs with Coulomb energy \cite{morf03}. 
This method has been previously used for gap estimations for the $\nu=5/2$ FQHS, e.g. Refs.\cite{morf03, gamma}.
According to this method, the measured energy gaps scale with the Coulomb energy according to $\Delta^{meas}=\delta^{int} E_C - \Gamma$,
where the term $\Gamma$ is commonly referred to as the disorder broadening parameter.
In the inset of Fig.4 we plot the experimental gaps of $\nu=4/11$ and $\nu=5/13$ FQHSs against Coulomb energy. 
Fitting these two points, we extract the intrinsic gap $\delta^{4/11} \approx 0.0028$, which is of the same order of magnitude as 0.002,
the value found in a recent numerical simulation\cite{mukherjee14}. The similarilty of the two intrinsic gaps, the one extracted from our data and that
from Ref.\cite{mukherjee14}, supports the idea of novel topological order of the WYQ type in the FQHS at $\nu=4/11$.

So far we have established incompressibility, i.e. the formation of a gap, 
at $\nu=4/11$ and the agreement of the measured and the calculated intrinsic gaps.
These  are necessary but not sufficient conditions for establishing the formation of WYQ state at this filling factor. 
Another necessary condition for the formation of WYQ state  is a full spin polarization of CFs at this filling factor.
Numerics cannot yet accurately predict in the thermodynamic limit the spin polarization field for realistic sample parameters 
\cite{goerbig,park,chang2,chang,AWojs04,mukherjee14,mukherjee15}.
Experimentally the tilted field technique may reveal the spin state of the FQHS at $\nu=4/11$ \cite{WPan03}. However,
because of the weakness of this state, convincing evidence of a spin transition remains an extremely challanging task.

In contrast to the temperature dependence of $R_{xx}$ at $\nu=4/11$ and $5/13$, that at
$\nu=3/8$ is very different. 
While the depression in $R_{xx}$ is present at  at $\nu=3/8$ to the lowest temperatures, its magnitude
does not seem to develop significantly as the temperature is dropped.
Furthermore, as seen in Fig.3, $R_{xx}$ at $\nu=3/8$  continues to increase
monotonically with a decreasing $T$.
Such an increase is inconsistent with an activated behavior and incompressibility
despite the existence of a depression in $R_{xx}$ at this filling factor. In the absence of incompressibility, the 
ground state at $\nu=3/8$ and, as discussed earlier, at $6/17$ are not accertained to be a FQHS.

 A FQHS at $\nu=2+3/8$ 
 has been observed in several samples \cite{WPan04,cell,choi,kumar10,deng1,pan12,deng2} 
 and it is believed that this state,  similarly to the $\nu=5/2$ FQHS, is a paired CF state \cite{toke08}.
 However, the formation of a FQHS at $\nu=2+3/8$ in the second Landau level, does not guarantee the formation of a 
 FQHS at $\nu=3/8$ in the lowest Landau level \cite{shibata04,shibata05,scarola,mukherjee12}. 
 It is interesting to compare other two filling factors to the half-filled Landau levels in the lowest and second Landau level.
 At $\nu=2+1/2=5/2$ a FQHS develops \cite{willett}, whereas at $\nu=1/2$ there is a compressible Fermi sea of CFs \cite{du,kang}.
 For these pairs of filling factors, the difference in the interactions between the CFs in different Landau levels clearly
 leads to very different ground states. It is thus not surprising that the presence of a FQHS at $\nu=2+3/8$ may be accompanied by
 a compressible state at $\nu=3/8$. Generally speaking,
the conditions of the formation  of paired CF states are currently not understood.
 Even though we do not observe incompressibility at $\nu=3/8$, the possibility of opening of a gap at 
 this filling factor in either higher quality samples or at lower temperatures can not be ruled out at this stage.
 
 In conclusion, because of the unusual interactions between the composite fermions, several fractional quantum Hall states 
 in the filling factor range $1/3<\nu<2/5$ are expected to have novel topological order. We performed
 an ultra-low temperature study of this region. We presented compelling evidence that the ground states at $\nu=4/11$ and $5/13$ 
 are incompressible. Our observations ensure genuine fractional quantum Hall ground states and pave the way towards establishing
 exotic topological order at these two filling factors. 
 In contrast, the ground state in our sample at two other filling factors in this region,
 namely at $\nu=3/8$ and $6/17$, are compressible to the lowest temperatures accessed. The development of 
 fractional quantum Hall states at the latter two filling factors remains an open question.

{\it Note added.} Recently, we became aware of a work by Pan
et al. \cite{Pan14} reporting on magnetoresistance at similar
Landau level filling factors.

\begin{acknowledgments}
N.S. and G.A.C. acknowledge funds from the NSF grant DMR-1207375. The work at Princeton was funded by the Gordon and Betty Moore Foundation through Grant 
GBMF 4420, and by the National Science Foundation MRSEC at the Princeton Center for
Complex Materials. We thank J. Jain for useful discussions.
\end{acknowledgments}


\end{document}